\documentclass[12pt]{article}

\usepackage{graphicx}
\usepackage{amsmath}
\usepackage{amssymb}
\usepackage{enumerate}
\usepackage{verbatim}
\usepackage{natbib}
\usepackage{longtable}

\newcommand{\Jyvaskyla}{Jyv{\"a}skyl{\"a}}

\begin{document}
\title{Survey data and Bayesian analysis: a cost-efficient way to estimate customer equity}
\author{Juha Karvanen$^1$*, Ari Rantanen$^2$ and Lasse Luoma$^3$\\
\\
$^1$ Department of Mathematics and Statistics,\\
University of \Jyvaskyla,\\
\Jyvaskyla, Finland\\
juha.t.karvanen@jyu.fi\\
*corresponding author\\
\\
$^2$ Sanoma Media Finland,\\
Helsinki, Finland\\
\\
$^3$ Tietoykk{\"o}nen Oy,\\
\Jyvaskyla, Finland
}

\maketitle


\section*{Abstract}
We present a Bayesian framework for estimating  the customer lifetime value (CLV) and the customer equity (CE) based on the purchasing behavior deducible from the market surveys on customer purchasing behavior. The proposed framework systematically addresses the challenges faced when the future value of customers is estimated based on survey data. The scarcity of the survey data and the sampling variance are countered by utilizing the prior information and quantifying the uncertainty of the CE and CLV estimates by posterior distributions. Furthermore, information on the purchase behavior of the customers of competitors available  in the survey data is integrated to the framework. The introduced approach is directly applicable in the domains where a customer relationship can be thought to be monogamous.

As an example on the use of the framework, we analyze a consumer survey on mobile phones carried out in Finland in February 2013. The survey data contains consumer given information on the current and previous brand of the phone and the times of the last two purchases. 

~\\
\noindent Keywords: Bayesian estimation, Brand switching, Customer equity, Customer lifetime value, Survey


\section{Introduction}

The monetary value of a future relationship with a customer is a fundamental concept for the rational long-term management of the customer base and the planning of marketing activities. In recent years, a lot of effort has been invested to develop models for estimating the future value of a customer relationship, or {\em customer lifetime value} (CLV)~\citep{BejouReview06,GuptaLehmannReview05,Blattberg08}. Thanks to this development, companies are nowadays able to utilize forward looking monetary estimates when assessing the value of the company~\citep{Bauer03,Pfeifer11}, planning marketing actions~\citep{Kumar08} or optimizing the customer base and allocating their resources~\citep{VenkatesanKumar04, KumarPetersen05, rust2001driving}.  If applied properly, CLV and its direct derivative, the future value of the total customer base, or {\em customer equity} (CE)~\citep{KumarGeorge07},  are able to reveal customers that are most valuable to a company in the long run, and direct the actions to produce optimal
return on marketing investments.


For companies with an access to customer level purchase histories, CLV can be estimated for each customer, at the {\em individual level}, based on the data on the past purchases of an individual customer and stochastic modeling of the general purchasing behavior of the population~\citep{Schmittlein87, Fader05, Fader05b}. For many companies, however, purchase histories of individual customers are not available and CLV models based on such data are not applicable. This can be due to many reasons. Many companies producing consumer goods or services use retailers to sell their products and thus lack the direct interaction with their end customers. In some cases, the time interval between repurchases is long, and gathering meaningful individual level purchase histories takes a lot of time. And even in the cases where the individual level customer data would be in principle available, the lack of maturity of company's data gathering process, the strict privacy policies or the prohibitive cost of collecting purchase data might not allow the utilization of the data for involved statistical modeling.

As all companies need to plan their marketing actions and  portfolio to maximize return on investment (ROI), also companies without an access to individual purchase histories would benefit from understanding the average CLV of different customer segments. 
For example, it is very common that the company manufacturing the consumer goods is much bigger than individual retailers that distribute the goods to consumers. Hence the manufacturing company needs to make substantial marketing investments to ensure the  demand of its products, even if the customer relationship, and data on the individual purchase histories, is owned by the retailers. Ensuring that these marketing investments are optimally distributed is a major task for many large manufacturing companies. The goal of the paper is to present a practical method for the estimation of CLV and CE when company level data on individual purchase histories are not available. Below we introduce a modeling framework to estimate CLV and CE from the survey data. The use of the data from a market survey solves the problem of data availability and may be a significantly cheaper option than implementing the full scale collection of individual purchase histories. 

Survey data have both advantages and disadvantages compared to customer registries. The data from the customer registry may suffer from adverse selection due to the cohort heterogeneity \citep{FaderHardie10} and thus give a biased view on retention. Survey data are drawn as a random sample from the whole population and therefore provide unbiased estimates of retention and churn, even in a non-contractual business setting. In addition, in a survey the respondents can be asked also on their purchase intentions, which are not visible from the purchase histories. In an anonymous survey, respondents can also be asked for rich background information that can be utilized to segment the customer base. Collecting the same information for the total customer base can be considered intrusive.




On the other hand, collecting information on the customer purchasing behavior via surveys has limitations that need to be addressed when CLV is estimated based on survey data. Compared to the full purchase histories, survey data are scarce. Survey respondents can be realistically assumed to remember only few latest transactions. Information on the timing of the transactions is also imprecise, making the purchase data interval censored. And even when the survey respondents form a representative sample of the whole customer base, fairly limited number of observations prohibits the use of complex statistical models, as they would easily overfit to the limited training data and produce biased forecasts on CLV. Limited number of sample observations also introduce a sampling variance to the CLV and CE estimates absent in data describing the full purchase histories of individual customers.

Our framework simultaneously addresses the listed challenges and can make the full use of information available in the survey data, including customer behavior with competitors. The framework uses survey data to model the brand switching behavior of customers between the focal brand and its competitors. The scarcity of the survey data is compensated by the use of prior information to guide the estimation process. Bayesian analysis \citep{Gelman13,bayesmarketing} 
provides a natural way to handle the uncertainty of the prior information and the data. Bayesian models can directly deal with interval censored purchase data.  The estimated Bayesian model not only gives the point estimates of the CLV and CE of different customer segments, but also quantifies the uncertainty of the estimates. This greatly increases the potential use of the estimates in the real world decision making, where the risk, or the expected opportunity loss, from the wrong marketing decision needs to be understood \citep{Hubbard10}.

Despite of its flexibility and usefulness, the Bayesian approach has not been frequently applied to the CLV and CE estimation. \citet{borle2008customer} and \citet{abe2009counting} proposed hierarchical Bayes extensions to the Pareto/NBD model \citep{Schmittlein87}. \citet{nagano2013nonparametric} extended the model further by allowing more flexibility for the individual level regression parameters. \citet{jen2009importance} used hierarchical Bayes models to model the temporal dependence of purchase quantity and timing. \citet{singh2009generalized} used Markov Chain Monte Carlo (MCMC) based data augmentation framework for estimating CLV in a noncontractual context. Our framework differs from these works by focusing on survey data and the uncertainty of the recorded purchase times. The cost-efficiency perspective has not been earlier used to motivate the use of survey data and Bayesian analysis.  

The general Bayesian framework for the estimation of CLV and CE from survey data is presented in Section~\ref{sec:Bayes}. In Section~\ref{sec:survey}, we demonstrate the proposed framework for estimating CE from survey data by analyzing a survey with 536 respondents carried out in Finland in February 2013. The respondents provided information on the brand of their current and previous mobile phone and the times of the last two purchases. For each individual, we model the intensity of the purchase process and the personal probability of repurchase as latent variables that depend on the covariates measured in the survey. Conclusions are given in Section~\ref{sec:discussion}.




\section{Bayesian estimation of customer equity from survey data} \label{sec:Bayes}
The business objective is to estimate the CLV and CE of the major companies for mobile phones or other electronic devices in a geographically specified market. Assume that the data on the purchase behavior are obtained for a small random sample of the population. The collected data contains answers to the following questions:
\begin{enumerate}
 \item What is the brand of your current device?
 \item When did you buy your current device?
 \item What was the brand of your previous device?
 \item When did you buy your previous device?
 \item Which brand would be the most interesting for you if you were to buy a new device now?
\end{enumerate}
In addition, there may be various questions on the background of the customer, such as age and gender, which can be used to define the customer segments. 

\begin{figure}[h]
  \includegraphics[width=\columnwidth]{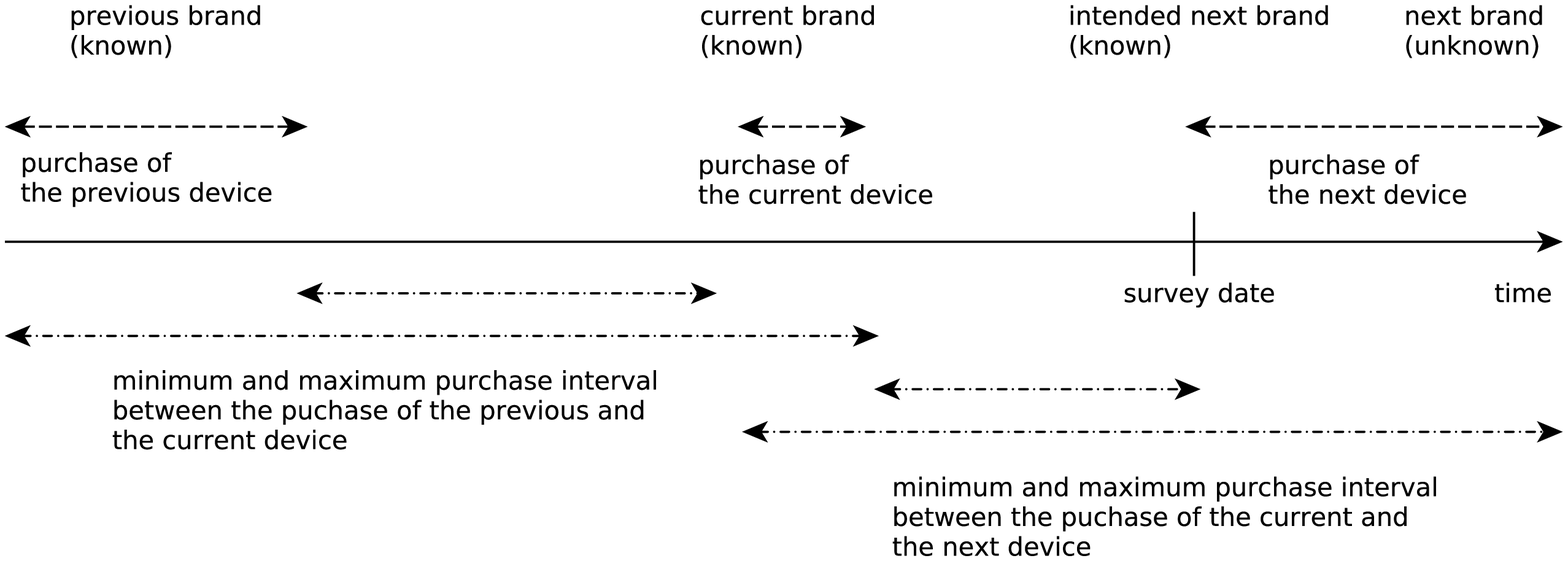}
  \caption{Illustration of the interval censoring for the purchase intervals. In `Historical repurchase' model, the data on the current brand and the previous brand are used; In `Intended repurchase' model, also the data on the intended next brand is used.}\label{fig:timeline}
\end{figure}

The structure of the data is illustrated in Figure~\ref{fig:timeline}. The purchase interval $T_i$ is defined as the difference of the dates from questions 2 and 4. However, the survey respondents do not usually remember these dates exactly, which leads to interval censored dates and consequently to interval censored purchase intervals. It follows that the models and methods used in the analysis should be capable for handling interval censored data. This condition is fulfilled by Bayesian analysis in a natural way whereas many frequentist methods would require special techniques to deal with interval censoring. 

It is also possible to utilize the data on the intended next brand (question 5). In `Intended repurchase' model, it is assumed that the next brand will be the intended brand. The time of the next purchase is unknown but the difference between the survey date and the purchase date of the current device (question 2) can be used  as a lower limit for the next purchase interval. The obtained right censored purchase intervals are referred as backward recurrence times and are potentially problematic in the maximum likelihood estimation \citep{Allison:survival}. Here the problems are avoided because the right censored times are used together with the interval censored times from questions 2 and 4. The interval censored times provide information on the right tail distribution of the purchase intervals which cannot be learned from right censored times. The comparison between `Intended repurchase' model and `Historical repurchase' model where customer intentions are not utilized, may reveal the direction of the future changes in the brand popularity.

The general procedure for Bayesian estimation of CE with survey data has the following steps:
\begin{enumerate}
 \item Specify a model for the customer behavior and choose the prior distributions for the model parameters. 
 \item Design the survey so that the model parameters can be efficiently estimated and collect the survey data
 \item Use Markov chain Monte Carlo (MCMC) or other simulation techniques to generate observations from the joint posterior distribution of the parameters. 
  \item For each set of parameters generated from the joint posterior distribution and for the each individual in the sample, generate several purchase histories for the forthcoming years. From these purchase histories, calculate the individual CLVs and the CE as their sum. These values are observations from the posterior distribution of the CE of the survey sample.
  \item Using the knowledge on the size of the market, scale the CE distribution of the sample to present the whole customer population.
\end{enumerate}
The procedure can be applied with a wide variety of different models. On one hand, the model needs to capture the purchase behavior of customers with a reasonable accuracy. On the other hand, the model needs to be identifiable from the (limited) survey data and general enough to avoid overfitting. 

In Section~\ref{sec:survey} we demonstrate how the general approach can be applied to solve a real-world problem of estimating the CLV and CE for the major mobile handset makers in Finnish market. The example shows how the model for the customer behavior can be tailored  to the business domain and available survey data. In Appendix we empirically validate the proposed approach with a general semi-Markov brand switching model and simulated data. The example in Appendix shows that the total CE can be estimated with a sufficient accuracy from survey samples of realistic size.




\section{Customer equity for mobile phone brands} \label{sec:survey}
\subsection{Survey data} \label{sec:data}


To demonstrate the practical applicability of the proposed framework, we apply it to estimate the CLV of different mobile phone brand in Finland from the real survey data. The mobile phone data were collected in February 2013 together with the National Consumer Net Shopping Study conducted by market research company Tietoykk{\"o}nen Oy. The target group was 15--79 years old mobile phone owners in Finland. The data collection method was telephone interviews by using a computer-assisted telephone interviewing (CATI) system. The sample source was targeting service Fonecta Finder B2C, which contains all publicly available phone numbers in Finland. Random sampling was made by setting quotas in respondents’ gender, age and region in the major region level excluding {\AA}land autonomic region. The sample size was 536 completed interviews. Table~\ref{tab:surveydata} shows the structure of the data. Compared to Finnish official statistics \citep{yearbook} the amount of women and youngest age group (15--24 years) in the data is slightly too small and men and oldest age group (65--79 years) too large.

\begin{center}
\begin{longtable}{llccc}
\caption{Respondents of the mobile phone survey by gender, age and major region} \label{tab:surveydata}\\
& & &  & Proportion in  \\
 & &   & Proportion &  population  \\
 & & Sample size & in sample & (age 15--79 y) \\
\hline
Gender & Man & 285 & 53\% & 50\% \\
& Woman & 251 & 47\% & 50\% \\
\hline
Age & 15--24 years & 63 & 12\% & 16\% \\
& 25--34 years & 88 & 16\% & 16\% \\
& 35--44 years & 73 & 14\% & 16\% \\
& 45--54 years & 89 & 17\% & 18\% \\
& 55--64 years & 103 & 19\% & 18\% \\
& 65--79 years & 120 & 22\% & 17\% \\
\hline
Region & Helsinki-Uusimaa & 143 & 27\% & 29\% \\
& Southern  & 91 & 17\% & 22\% \\
& Western  & 155 & 29\% & 27\% \\
& Northern and Eastern  & 147 & 27\% & 23\% \\
\hline
Total & & 536 & 100\% & 100\%
\end{longtable}
\end{center}

Finnish market is well saturated with mobile phones. Practically all household have at least one mobile phone at their disposal. In total, there were 9.3 million active mobile subscriptions at the end of year 2012~\citep{FICORA2012}, when the size of the Finnish population was 5.4 million at the same time.

Mobile operators sell the majority of mobile phones. While some operators offer subsidies if the device is purchased together with an operator contract, the subsidies are not very aggressive. Furthermore, devices are not locked to a contract or to an operator.  All major operator offer all major device brands. Hence, Finnish mobile device market is not as operator controlled as  US market, for instance, and a consumer is not heavily directed by an operator to choose a certain device brand or a device repurchase rate. 

All 536 survey respondents had a mobile phone. The respondents answered the following questions (originally in Finnish):
\begin{enumerate}
 \item What is the brand of your mobile phone?
 \item When did you purchase your mobile phone? (year and month; if the month was not recalled the season was asked)
 \item What was the brand of your previous mobile phone?
 \item When did you purchase your previous mobile phone? (year and month; if the month was not recalled the season was asked)
 \item Which brand would be the most interesting for you if you were to buy a mobile phone now?
 \item Is your mobile phone a smart phone, a feature phone with an internet connection or a phone without an internet connection?
\end{enumerate}
In addition, the respondents where asked for their gender, age group (six categories: 15--24, 25--34, 35--44, 45--55, 55--64 and 65--79 years), geographical region (Helsinki-Uusimaa, Southern Finland, Western Finland, Northern \& Eastern Finland) and income of the household (five categories: 30,000 euros or less, 30,001--50,000 euros, 50,001--70,000 euros, over 70,000 euros and no answer).

Table~\ref{tab:agebrand} presents the distribution of the brand by gender and age. It can be seen that there is a clear association between the age group and the current brand. 73\% of the respondents have Nokia as their current phone but Nokia's share of the installed base varies from the 93\% of the age group 65--79 years to the 46 \% of the age group 15--24 years. In contrast, Apple and Samsung are relatively strong in the younger age groups. This  suggest that the CLV analysis should be stratified by the age group.

\begin{center}
\begin{longtable}{llcccc}
\caption{Brand of current mobile phone by gender and age} \label{tab:agebrand}\\
& & \multicolumn{4}{c}{Current brand} \\
& & Nokia & Apple & Samsung & Others, n/a \\
\hline
Gender & Man & 70\% & 6\% & 18\% & 6\% \\
& Woman & 76\% & 9\% & 11\% & 4\% \\
\hline
Age & 15--24 years & 46\% & 14\% & 27\% & 13\% \\
& 25--34 years & 52\% & 16\% & 26\% & 6\% \\
& 35--44 years & 71\% & 3\% & 22\% & 4\% \\
& 45--54 years & 78\% & 6\% & 13\% & 3\% \\
& 55--64 years & 81\% & 8\% & 8\% & 4\% \\
& 65--79 years & 93\% & 2\% & 3\% & 3\% \\
\hline
Total & & 73\% & 7\% & 15\% & 5\%
\end{longtable}
\end{center}

The purchase times are interval censored: the respondents are asked only for the purchase month, not for the day and many respondents could not recall the time of the purchase. Out of 536 respondents, 310 were able to tell the purchase month and year, additional 115 were able to tell the season and year, 74 mentioned only the year and 37 were not able to tell even the year. 19 respondents told that they did not have a mobile phone earlier or they cannot remember the brand. Out of 517 respondents who mentioned the previous brand, 117 were able to tell the purchase month and year, additional 91 were able to tell the season and the year, 146 mentioned only the year and 163 were not able to tell even the year. Figure~\ref{fig:timeline} illustrates the calculation of the minimum and the maximum purchase intervals. The maximum purchase interval of 200 months is assumed when the purchase year is missing.

The survey data allows the modeling of purchase intervals and brand choices but do not provide information on sell-in prices that are also needed to estimate CE.  Companies may obtain detailed data on pricing from their internal sources but in this example we rely on external information that is publicly available.
Information on the sell-in prices is collected from the quarterly reports of Nokia, Apple and Samsung. For the 4th quarter of 2012, Nokia reported average sales price (ASP) 186 euros for smart phones and 31 euros for (other) mobile phones . For the same period, the ASP for Apple was 641 US dollars (473 euros) and the ASP for Samsung 178 euros. As the actual ASPs for Finland have not been published, these global numbers are used in the CLV estimation. The Nokia ASP is calculated to be 68 euros using the ratio of smart phones and mobile phones for the Nokia owners in the data.

\subsection{Bayesian modeling}
\begin{figure}
  \includegraphics[width=\columnwidth]{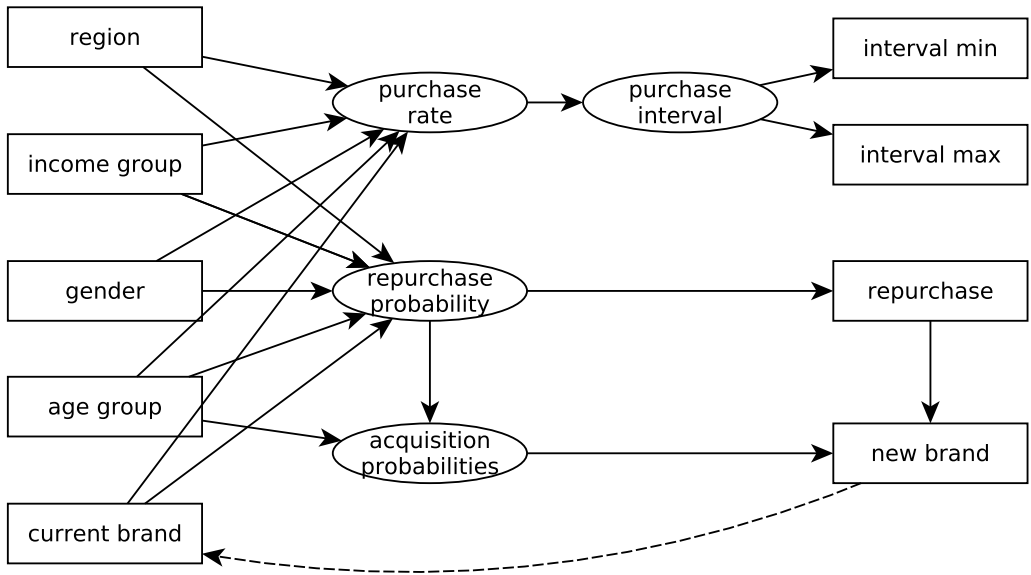}
  \caption{Graphical model for the data generating mechanism of the mobile phone survey. The observed variables are presented as rectangles and the latent variables as ellipses. The arrows describe the relationships between the variables. The dashed arrow from new brand to current brand indicates that the new brand will be the current brand when the next transaction is considered.}\label{fig:bayesmodel}
\end{figure}

The structure of the Bayesian model and the survey data is presented in Figure~\ref{fig:bayesmodel}. The observed covariates in the model are the age group, the income group, gender, the region and the current brand. The current brand is a dynamic covariate that changes at each transaction and the other covariates are assumed to be static. The latent variables in the model are the purchase rate $\lambda_i$, the repurchase probability $p_i$ and the acquisition probabilities for each brand. The purchase rate and the repurchase probability depend on all the covariates but the acquisition probabilities depend only on the age group because the small size of the data does not allow the estimation of acquisition probabilities for all covariate combinations. 
The purchase interval $T_i$ is not observed exactly but is interval censored with observed lower limit $t_i^{\textrm{min}}$ and upper limit $t_i^{\textrm{max}}$. The repurchase indicator $R_i \in \{0,1\}$ and the purchased brand are observed. The common covariates, region, income group, gender and age group control the dependence structure between the purchase intervals and the repurchase probabilities.

The model consists of three parts: the submodel for the purchase intervals, the submodel for the repurchase and the submodel for the choice of the new brand in the case of churn. The submodel for the purchase intervals can be presented as follows:
\begin{align*}
& T_i \in [t^{\textrm{min}}_i,t^{\textrm{max}}_i], \\
& T_i \sim \textrm{Gamma}(\kappa,\lambda_i), \\
& \kappa \sim \textrm{Gamma}(1,1), \\
& \log(\lambda_i) = \beta_0+\beta_{\textrm{agegr}[i]}+\beta_{\textrm{incomegr}[i]}+\beta_{\textrm{gender}[i]}+\beta_{\textrm{region}[i]}+\beta_{\textrm{brand}[i]},\\
& \beta_0 \sim N(0,1000),\;\;\beta_{\textrm{agegr}[i]} \sim N(0,1000),\;\;\beta_{\textrm{incomegr}[i]} \sim N(0,1000), \\
& \beta_{\textrm{gender}[i]} \sim N(0,1000),\;\;\beta_{\textrm{region}[i]}\sim N(0,1000),\beta_{\textrm{brand}[i]} \sim N(0,1000).
\end{align*}
The purchase intervals are assumed to follow the Gamma distribution with shape parameter $\kappa$ common for all intervals and scale parameter $\lambda_i$ defined separately for each interval. Assuming the Gamma distribution here increases the flexibility of the model compared to the exponential distribution. An informative Gamma prior is assumed $\kappa$ because the value of the shape parameter is not expected to be very far from $\kappa=1$, which corresponds to the exponential distribution. The scale parameter $\lambda_i$ follows a log-linear model where the regression coefficients for the covariates have uninformative priors. Only main effects are included because the small size of the data does not allow the interactions of the covariates to be estimated reliably. The regression coefficients are defined so that the reference is a poor young man who lives in Helsinki-Uusimaa and owns a Nokia phone. The parameter $\beta_0$ defines the value of the scale parameter for the reference. 

The submodel for the repurchase can be presented as follows:
\begin{align*}
& R_i \in \{0,1\},\\
& \textrm{logit}(R_i=1) = \alpha_0+\alpha_{\textrm{agegr}[i]}+\alpha_{\textrm{incomegr}[i]}+\alpha_{\textrm{gender}[i]}+\alpha_{\textrm{region}[i]}+\alpha_{\textrm{brand}[i]},\\
& \alpha_0 \sim N(0,1000),\;\;\alpha_{\textrm{agegr}[i]} \sim N(0,1000),\;\;\alpha_{\textrm{incomegr}[i]} \sim N(0,1000),\\
& \alpha_{\textrm{gender}[i]} \sim N(0,1000),\;\;\alpha_{\textrm{region}[i]} \sim N(0,1000),\;\;\alpha_{\textrm{brand}[i]} \sim N(0,1000).
\end{align*}
The repurchase indicator follows a logistic regression model where the regression coefficients for the covariates have uninformative priors. The reference is defined as above.

Finally, the submodel for the choice of the new brand in the case of churn can be presented as follows:
\begin{align*}
& P(\textrm{newbrand}[i]=v \mid \textrm{brand}[i]=u, v \neq u, \textrm{agegr}[i]=h, R_i=0)= \frac{w_{vh}}{\sum_{\{j:j \neq u\}} w_{jh}},\\
& w_{jh} \sim \textrm{Beta}(2,2) \textrm{ for all } j,h. 
\end{align*}
The weights $w_{jh}$, where $j$ refers to the brand and the $h$ refers to the age group, are used as auxiliary variables that describe the relative popularity of each brand. The probability of a brand to be selected equals the relative popularity of the brand divided by the sum of the relative probability of all brands except the current. The use of the $\textrm{Beta}(2,2)$ prior for the weights instead of the uniform prior stabilizes the scale of the weights and ensures the convergence in the MCMC estimation. With these definitions, the choice of the new brand follows multinominal distribution where the selection probabilities depend on the age group. 

The data can be utilized in the estimation in two alternative ways: The model `Historical repurchase' uses only the actual historical purchase events whereas the model `Intended repurchase' assumes that the next brand will be the brand the individual is most interested in according to the survey (question 5 in the list of questions above). These two alternatives are also illustrated in Figure~\ref{fig:timeline}. The purchase intervals are defined and analyzed as described in Section~\ref{sec:Bayes}. The same distributional assumptions are used for both the models.

The estimation is carried out using OpenBUGS 3.2.2 \citep{openbugs}, R \citep{R} and R2OpenBUGS R package \citep{R2OpenBUGS}. The BUGS code is provided in Appendix~2. The convergence of the MCMC chains is monitored separately for each parameter using the interval criterion proposed by \citet{brooksgelman}. 

\subsection{Estimated model}
The estimated parameters are presented in Table~\ref{tab:estimates} for the models `Historical repurchase' and `Intended repurchase'. Theoretically the parameters for the submodel for the purchase intervals are the same and the small differences are only due to the MCMC estimation. The differences in the parameter estimates in the submodels for the repurchase and the brand choice reflect real differences between the models `Historical repurchase' and `Intended repurchase'. The shape parameter of the distribution of the purchase intervals has value $\kappa=1.5$, which indicates that the hazard of transaction increases as the time from the last purchase increases. For the exponential distribution $\kappa=1$, the hazard is constant. The estimated average purchase interval is 2.9 years based on simulations from $\textrm{Gamma}(\kappa,\lambda_i)$.  For Apple and Samsung the purchase rates are higher compared to Nokia (parameters $\beta_\textrm{Apple}$ and $\beta_\textrm{Samsung}$). There are no clear differences in the purchase rates between the age groups, gender, income group or geographical areas (parameters from $\beta_\textrm{25--34}$ to  $\beta_\textrm{Eastern \& Northern}$). 

When repurchase probabilities are considered, the owners of Apple or Nokia seem to more loyal than the owners of Samsung or other brands (parameters $\alpha_\textrm{Apple}$, $\alpha_\textrm{Samsung}$ and $\alpha_\textrm{Other brand}$). The older age groups are more loyal to their current brand than the younger age groups. Gender, income group or geographical region do not have a major effect to the repurchase probabilities. When the models `Historical repurchase' and `Intended repurchase' are compared, it can be seen that in the latter model the repurchase probabilities are higher for all brands but especially for Apple and Samsung.   

\begin{center}
\begin{longtable}{lcccc}
\caption{Estimated model parameters for the mobile phone survey. The mean and 95\% credible interval of the posterior distribution are presented. The reference categories with the regression coefficients fixed to zero are: Nokia, 15--24 years, Men, below 30k euros and Helsinki-Uusimaa.} \label{tab:estimates}\\
 & \multicolumn{2}{c}{Historical repurchase} & \multicolumn{2}{c}{Intended repurchase} \\
Parameter & Mean & 95\% CI & Mean & 95\% CI\\ \hline
 $\kappa$  &   1.53  &  (1.40, 1.67)  &   1.54  &  (1.40, 1.68) \\ 
 $\beta_0$  &  -0.46  &  (-0.70, -0.22)  &  -0.46  &  (-0.72, -0.22) \\ 
 $\beta_\textrm{Apple}$  &   0.71  &  (0.46, 0.95)  &   0.71  &  (0.47, 0.94) \\ 
 $\beta_\textrm{Samsung}$  &   0.55  &  (0.37, 0.72)  &   0.55  &  (0.37, 0.72) \\ 
 $\beta_\textrm{Other brand}$  &   0.10  &  (-0.16, 0.35)  &   0.10  &  (-0.17, 0.34) \\ 
 $\beta_\textrm{25--34}$  &  -0.09  &  (-0.30, 0.13)  &  -0.08  &  (-0.30, 0.13) \\ 
 $\beta_\textrm{35--44}$  &  -0.23  &  (-0.45, 0.00)  &  -0.23  &  (-0.45, 0.00) \\ 
 $\beta_\textrm{45--54}$  &  -0.34  &  (-0.57, -0.12)  &  -0.34  &  (-0.56, -0.12) \\ 
 $\beta_\textrm{55--64}$  &  -0.44  &  (-0.65, -0.23)  &  -0.44  &  (-0.65, -0.22) \\ 
 $\beta_\textrm{65--79}$  &  -0.62  &  (-0.83, -0.40)  &  -0.62  &  (-0.83, -0.39) \\ 
 $\beta_\textrm{Woman}$  &   0.06  &  (-0.05, 0.17)  &   0.06  &  (-0.05, 0.17) \\ 
 $\beta_\textrm{30k--50k EUR}$  &   0.20  &  (0.04, 0.35)  &   0.19  &  (0.04, 0.35) \\ 
 $\beta_\textrm{50k--70k EUR}$  &   0.16  &  (-0.02, 0.33)  &   0.16  &  (-0.02, 0.33) \\ 
 $\beta_\textrm{70k- EUR}$  &   0.22  &  (0.02, 0.42)  &   0.22  &  (0.02, 0.41) \\ 
 $\beta_\textrm{unknown income}$  &   0.28  &  (0.09, 0.47)  &   0.28  &  (0.09, 0.47) \\ 
 $\beta_\textrm{Southern}$  &  -0.02  &  (-0.19, 0.15)  &  -0.02  &  (-0.19, 0.15) \\ 
 $\beta_\textrm{Western}$  &  -0.16  &  (-0.31, -0.01)  &  -0.16  &  (-0.31, -0.01) \\ 
 $\beta_\textrm{Eastern \& Northern}$  &   0.01  &  (-0.14, 0.15)  &   0.00  &  (-0.15, 0.16) \\ 
 $\alpha_0$  &   0.25  &  (-0.59, 1.07)  &   0.63  &  (0.02, 1.26) \\ 
 $\alpha_\textrm{Apple}$  &  -0.05  &  (-1.53, 1.63)  &   0.87  &  (0.05, 1.79) \\ 
 $\alpha_\textrm{Samsung}$  &  -1.11  &  (-1.86, -0.35)  &  -0.65  &  (-1.11, -0.17) \\ 
 $\alpha_\textrm{Other brand}$  &  -3.11  &  (-4.08, -2.24)  &  -2.21  &  (-2.84, -1.58) \\ 
 $\alpha_\textrm{25--34}$  &   0.80  &  (0.05, 1.58)  &   0.94  &  (0.40, 1.49) \\ 
 $\alpha_\textrm{35--44}$  &   1.02  &  (0.22, 1.86)  &   0.78  &  (0.21, 1.35) \\ 
 $\alpha_\textrm{45--54}$  &   1.33  &  (0.52, 2.17)  &   1.25  &  (0.67, 1.83) \\ 
 $\alpha_\textrm{55--64}$  &   1.84  &  (1.05, 2.67)  &   1.44  &  (0.86, 2.01) \\ 
 $\alpha_\textrm{65--79}$  &   2.27  &  (1.47, 3.13)  &   1.43  &  (0.87, 1.99) \\ 
 $\alpha_\textrm{Woman}$  &   0.22  &  (-0.22, 0.66)  &   0.12  &  (-0.20, 0.43) \\ 
 $\alpha_\textrm{30k--50k EUR}$  &   0.42  &  (-0.19, 1.05)  &  -0.01  &  (-0.44, 0.42) \\ 
 $\alpha_\textrm{50k--70k EUR}$  &   0.06  &  (-0.64, 0.78)  &   0.05  &  (-0.47, 0.58) \\ 
 $\alpha_\textrm{70k- EUR}$  &  -0.60  &  (-1.37, 0.17)  &  -0.75  &  (-1.30, -0.20) \\ 
 $\alpha_\textrm{unknown income}$  &  -0.34  &  (-1.02, 0.35)  &  -0.24  &  (-0.74, 0.27) \\ 
 $\alpha_\textrm{Southern}$  &  -1.09  &  (-1.77, -0.41)  &  -0.75  &  (-1.24, -0.27) \\ 
 $\alpha_\textrm{Western}$  &  -0.16  &  (-0.78, 0.47)  &  -0.04  &  (-0.50, 0.41) \\ 
 $\alpha_\textrm{Eastern \& Northern}$  &  -0.33  &  (-0.93, 0.31)  &  -0.36  &  (-0.80, 0.08)
\end{longtable}
\end{center}

\subsection{Model diagnostics}
\begin{figure}
\includegraphics[width=\columnwidth]{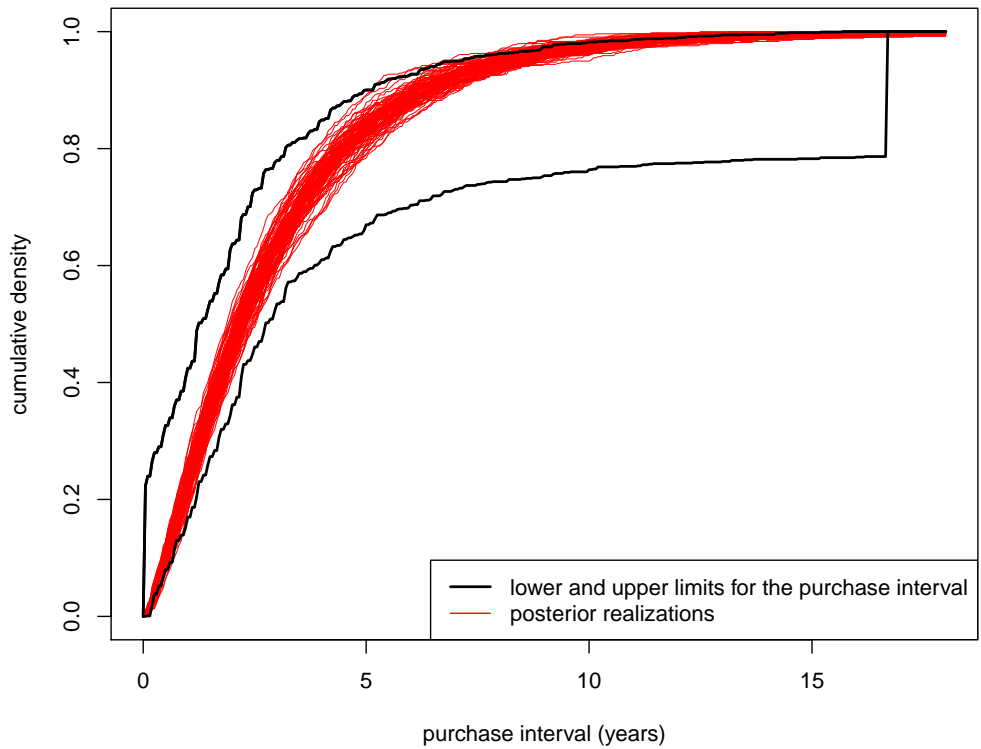}
  \caption{Cumulative distribution functions for simulated purchase intervals and the minimum and the maximum purchase intervals in the data. The jump  in the upper limit is due to upper limit of 200 months.}\label{fig:diagnostics}
\end{figure}

The model fit was studied by comparing the purchase intervals and repurchase probabilities between the data and the model. The real purchase intervals are not observed directly but are interval censored which complicates the checking of the model fit. For each individual in the data, the 9000 values of parameters $\kappa$ and $\lambda_i$ are drawn from their posterior distribution available as the result of the MCMC estimation. For each realization of parameters $\kappa$ and $\lambda_i$, purchase intervals are generated and 9000 posterior realizations of the cumulative distribution function (CDF) of the purchase interval are obtained.  One hundred randomly selected realizations of the CDF of the purchase interval are plotted in Figure~\ref{fig:diagnostics} together with the CDFs of the minimum and the maximum purchase intervals in the data. The CDF of the minimum (maximum) purchase intervals is obtained as the CDF of the lower (upper) limits of the purchase intervals. In general, the posterior CDFs lie between the CDFs of the minimum and the maximum purchase intervals, which indicates that the model fits to the data.

The fit of the estimated repurchase and acquisition probabilities is checked by regenerating the distribution of the current brand on the basis of the previous brand and the estimated posterior parameters and comparing this to the observed distribution of the current brand in the data. The comparison in Table~\ref{tab:diagnostics} shows that the frequencies in the data are always inside the 95\% credible intervals calculated from the simulations. We conclude that the model fits to the data.

\begin{center}
\begin{longtable}{lcl}
\caption{Frequencies for the current brand according to the data and the simulations (mean and the 95\% credible intervals). In the simulations, the model parameters  follow the posterior distributions and in the starting situation each individual has a phone with the previous brand indicated in the data.} \label{tab:diagnostics}\\
Brand & Freq. in data & Mean (95\% CI) in simulation\\ \hline
 Nokia  &  390  &  373.8    (356.5, 390.7) \\ 
 Apple  &  40  &  ~47.7    (37.1, 59.1) \\ 
 Samsung  &  80  &  ~79.0    (65.6, 92.8) \\ 
 Other  &  26  &  ~35.6    (26.0, 46.1) 
\end{longtable}
\end{center}

\subsection{Estimation of CLV and CE}
Using the estimated posterior distributions of the parameters, 2000 purchase histories were generated for each individual in the sample. Due to the volatile nature of the mobile phone markets, we restricted the simulation to cover only the next five years from March 2013 to February 2018. It can be argued that the uncertainty of the revenues in the future is so high that the decisions on the marketing actions should not be based on the transactions beyond the next five years. The each simulated purchase history contains all purchases with the purchase date and purchased brand from the next five years. As the data do not offer information on the customer level costs (calls to the customer care etc.), we exclude these cost and concentrate on the revenues. With the ASPs given in Section~\ref{sec:data} and the annual discount rate 10\%, the net present value the purchases can be calculated. 

As a result, we obtain 2000 simulated five-year CLVs for each individual. All calculations were repeated by using two alternative models and data: a) only historical purchases or b) both the historical and intended purchases. The average five-year CLV by the brand and the customer status are presented in Table~\ref{tab:averageclv}. As expected the CLV was higher for the current customers than non-customers. The average CLV of a current Nokia owner for Nokia is 70 euros, which roughly means that the customer is going to buy one Nokia phone during the next five years. The average value of a current Apple owner for Apple is 840 euros or 1183 euros depending on the model. The latter number is equivalent to the value of purchasing Apple every second year during the next five years. The result reflects the high purchase rate, the high repurchase probabilities and the high ASP for Apple. The average value of a current Samsung owner for Samsung is approximately equivalent to the value of buying one Samsung phone during the next five years. When the intentions are taken into account, the CLV of Apple and Samsung increase and the CLV of Nokia decreases.  The results also demonstrate the impact of survey uncertainty: the credible intervals of the mean are wider for Apple and Samsung than for Nokia because the number of purchases in the data are smaller. 

\begin{center}
\begin{longtable}{lccccc}
\caption{Average five-year CLV with its 95\% credible interval by the brand and the customer status. The values are in euros.} \label{tab:averageclv}\\
 & Current  & \multicolumn{2}{c}{Historical repurchase}   & \multicolumn{2}{c}{Intended repurchase} \\
 Brand     & customer & mean CLV  & 95\% CI  & mean CLV  & 95\% CI  \\
\hline
Nokia  &  Yes  &  71  &  (62,82)  &  70  &  (61,79) \\ 
 Nokia  &  No  &  50  &  (35,65)  &  40  &  (29,52) \\ 
 Apple  &  Yes  &  840  &  (411,1335)  &  1183  &  (845,1538) \\ 
 Apple  &  No  &  98  &  (58,146)  &  96  &  (61,138) \\ 
 Samsung  &  Yes  &  182  &  (116,261)  &  234  &  (170,307) \\ 
 Samsung  &  No  &  38  &  (25,53)  &  40  &  (29,53) 
 \end{longtable}
 \end{center}
 
\begin{figure}
 \includegraphics[width=\columnwidth]{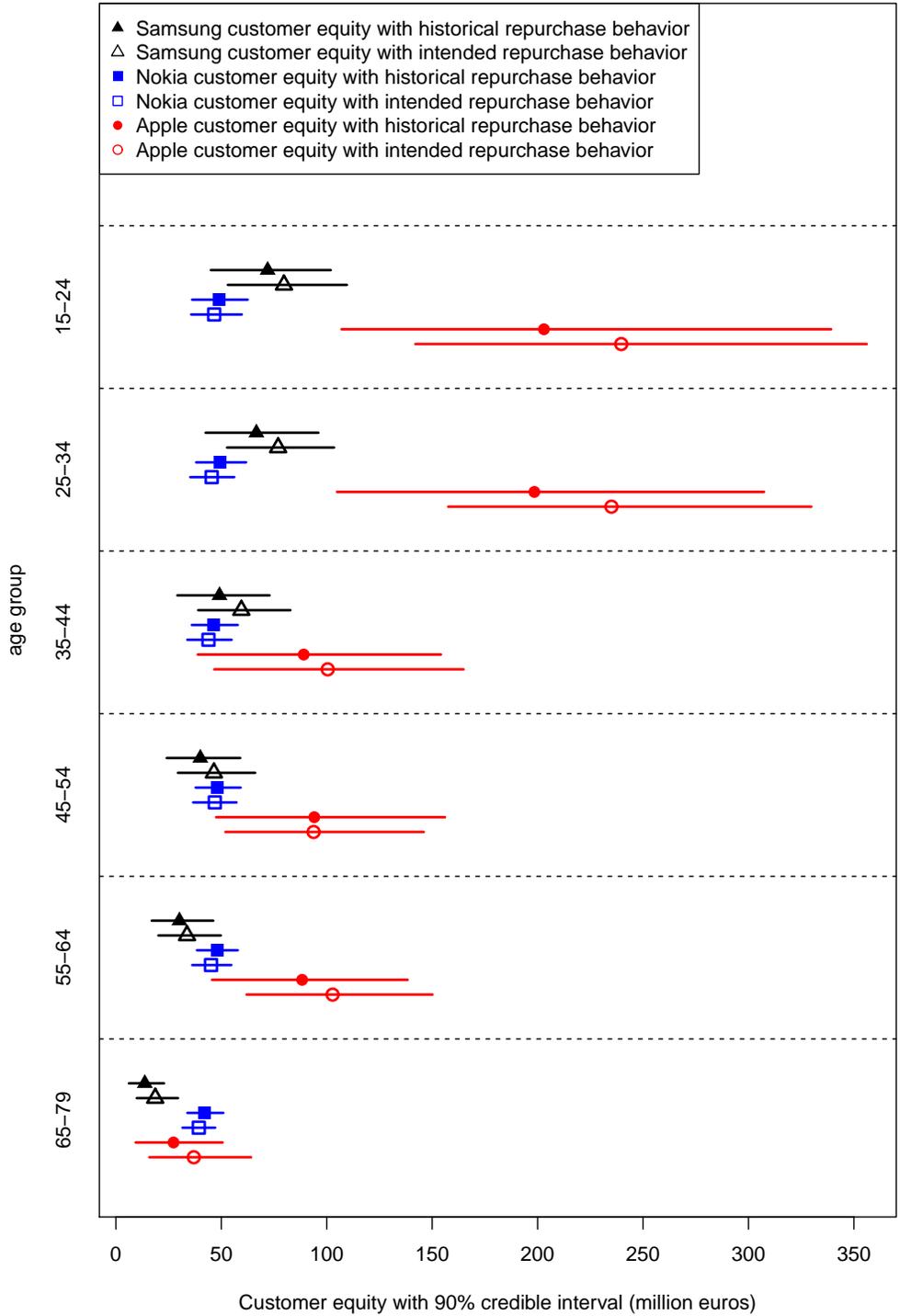}
  \caption{The customer equities by brand and age group calculated using the historical and the intended purchase behavior.}\label{fig:ce}
\end{figure}

The five-year CE by brand and age group are presented in Figure~\ref{fig:ce}. The five-year CE is calculated by scaling the average five-year CLV by the brand and the age group to the population level by using the population statistics by Statistics Finland given in Table~\ref{tab:surveydata}.  Apple has clearly the highest CE in Finland and the margin is largest in the young age groups. The survey uncertainty is reflected by the wide credible intervals for Apple.
Nokia and Samsung have approximately the same CE at the population level but Samsung is the strongest in the young age groups while the CE of Nokia comes equally from all age groups. The CE of Nokia can be estimated rather accurately from the survey data but considering Apple and Samsung a larger sample size would have been desirable.

\section{Discussion} \label{sec:discussion}
We have presented a cost-efficient Bayesian approach to estimate the average CLV and the CE on the basis of survey data. The presented approach is motivated by the need for the CLV based decision making in the absence of personal purchase histories. Most companies manufacturing consumer goods do not directly sell the goods to their end customer, and hence are unable to directly collect transactional purchase data at the level of individual customer. For these companies, surveys offer a natural option to obtain information about the purchasing behavior different customer segments.

Although the amount of the data is significantly smaller in surveys compared to customer registries, the survey based approach for the estimation of CE has important benefits. A survey gives insight also on the purchase behavior of the customers of the competitors. A properly collected sample represents the population and thus avoids the problems of cohort heterogeneity. Carrying out a survey is usually an easier option than organizing systematic collection of purchase histories and does not require investments to transactional data collection and storage. The use of survey data and Bayesian analysis could be a cost-efficient alternative to estimate CE for many companies that do not collect transactional data.

As always in survey sampling, the validity of the results depends on the representativeness of the sample and major differences in the survey response rates between the customer segments may bias the CE estimates. In many cases, the bias can be removed or reduced by stratified sampling or post-stratification which lead to unequal weighting of the individuals in the sample. Customer panels provide an alternative for cross-sectional surveys. Panels provide prospective data on the transactions but the population representativeness of the customer behavior may be a bigger problem than in surveys.

As a demonstration of the proposed survey based approach, we analyzed simulated data (results given in Appendix) as well as real data on the Finnish mobile markets. The applied statistical models were different in these examples to emphasize the generality of the approach. From the simulated data we learn that rather small sample sizes may be sufficient for unbiased  CLV and CE estimation. The same was observed in the real data example where reasonable CLV and CE estimates were obtained from a survey with only 536 respondents.


The example of mobile phone survey demonstrates the importance of high customer loyalty and high ASP over the the success in the past when forward-looking customer value is estimated. With the installed base of 73\% of the population, Nokia has expected five-year CE of 280 million Euros while Apple has expected five-year CE of 700 million Euros with the installed base of only 7\%.

The mobile phone market is known to be particularly volatile. Disruptive innovations, such as the introduction of Apple iPhone in 2007, are difficult to predict but may have a major impact to the future repurchase probabilities.  In fact, in September 2013 it was announced that the mobile phone business of Nokia will be acquired by Microsoft Corporation. This might or might not have an impact to consumer behavior in Finland.

Taking the unpredictability of the long-run market dynamics into account, the CE should not be understood as an attempt to forecast the market
development in the future but as a projection of the current state to the future. From this perspective, CLV and CE work as tools to compare brands and customer segments under the (unrealistic) assumption that the present model will describe also the future. The survey questions on the intended purchase
behavior provide insight on the validity of this assumption. Significant differences in the CE estimates with historical repurchase behavior and intended
repurchase behavior  indicate the unpredictability of the market dynamics. And even in the changing business landscape, some CLV and CE estimates are required for the decisions on the marketing actions. On the other hand, the differences between historical and intended behavior also reflect the actual changes in the market dynamics, as in the example case above. Hence, the proposed approach can be used to quantify the CE impact of an on-going change in the market in the short-term. 


An obvious weakness of the presented mobile phone survey example is the lack of individual level price data. The respondents could be naturally asked for the price paid for their current phone but this does not provide a straightforward solution to the problem. Operator subsidies and the recall bias complicate the data collection. Even in the best case the respondents can name only the retail price, not the sell-in price. Further, the purchase rate and the customer loyalty are aspects of individual level customer behavior whereas the phone prices in the future are not chosen by the customer but depend on the market. Despite the challenges, we recommend the price data to be collected in surveys.

We believe that the presented approach with survey data and Bayesian modeling will help in advocating the usefulness of the CLV and CE modeling even in the absence of personal purchase histories. Furthermore, we believe that the attention to the uncertainty of the CLV and CE will make the risks more explicit for the decision makers.

\section*{Appendix 1}

\subsection*{Model for the simulation example}
The estimation method is demonstrated with simulated data from a heterogeneous semi-Markov brand switching model which is defined as follows:
\begin{enumerate}
 \item The number of transactions made by an individual $i$ follows a Poisson process with the transaction rate $\lambda_i$.
\item Transaction rate $\lambda_i$ follows a Gamma distribution with the probability density function
\begin{equation*}
 f(\lambda_i \mid \gamma, \delta) = \frac{\delta^\gamma}{\Gamma(\gamma)} \lambda_i^{\gamma-1} e^{-\delta \lambda_i}, \quad \lambda_i>0,
\end{equation*}
where $\gamma>0$ is the shape parameter, $\delta>0$ is the rate parameter and $\Gamma$ stands for the Gamma function.
\item After any transaction, an individual may change the brand with a probability that depends on the current brand. For the focal company, the probability of repurchase for individual $i$ is $p_i$. The probability of acquisition, i.e. change from any competitor to the focal company, is $1-q_i$ for individual $i$.
\item Repurchase probability $p_i$ follows $\textrm{Beta}(\alpha_p,\beta_p)$ distribution and competitor repurchase probability $q_i$ follows $\textrm{Beta}(\alpha_q,\beta_q)$ distribution.
\end{enumerate}
The model is related to the long modeling tradition in the marketing literature \citep{herniter1971probablistic,schmittlein1985technical,vilcassim1991modeling,Fader05b}.
An individual is either in a state where she has made the last transaction with the focal company, or the individual is in a `competitor state', where she has made the last transaction with one of the competitors of the focal company. The transaction rate is assumed to be the same in the both states and independent on the transition probabilities.

Full purchases histories are generated for the population from where small survey samples are drawn. The CE estimated from the sample is compared with the CE of the population. The procedure is repeated for a number of survey samples to obtain information on the sampling variation.

The survey data collected for the individuals $i=1,2,\ldots,n$ are the current state $S^{(0)}_i$ (1 for the focal company and 0 for the competitors), the previous state $S^{(-1)}_i$, the time between the last two purchases $T_i$ and the time from the latest transaction $T^*_i$. As the transactions follow the Poisson process, the time between the purchases $T_i$ follows exponential distribution with rate $\lambda_i$. The time from the latest purchase to the day of the survey $T^*_i$ is an observation from the same exponential distribution because the Poisson process is memoryless.
For the current state it holds
\begin{equation}
 P(S^{(0)}_i=1)=p_i S^{(-1)}_i + (1-q_i) (1-S^{(-1)}_i).
\end{equation}
For the previous state it holds
\begin{equation}
 P(S^{(-1)}_i=1)=p_i S^{(-2)}_i + (1-q_i) (1-S^{(-2)}_i),
\end{equation}
where $S^{(-2)}_i$ is the state before the previous. For the state $S^{(-2)}_i$ there are no observations but the formula
\begin{equation}
 P(S^{(-2)}_i=1)=\frac{1-q_i}{2-q_i-p_i}
\end{equation}
follows from the equilibrium state of the Markov chain characterizing the brand switching.

\subsection*{Simulation setup}
We first simulate the complete purchase histories of 100,000 individuals who are divided between the focal company and the competitors according to the market shares. Then, by using the proposed approach, we estimate the average CLV and CE from the small `survey sample' of the simulated data, and compare the result to the true CLV and CE. 
The parameters used in the simulation are $\gamma=3$, $\delta=10$, $\alpha_p=4$, $\beta_p=6$, $\alpha_q=4$, $\beta_q=6$ and the intensity is defined as the number of transactions per year. The value of a purchase assumed to be 100 euros. The purchase histories are generated for the 40 years forward and 30 years backward from the time of the survey. The true CLVs and CEs are calculated using the whole population and the generated purchase histories for the forthcoming 40 years. With the annual discounting rate of 10 \% this leads to CE of 10.0 million euros for the population of the 100,000 individuals and an average CLV of 100 euros. For the current customers of the focal company, the average CLV equals 120 euros and for the current customers of competitors, the average CLV equals 91 euros. These numbers are
compared to the estimates from a small survey samples from the same population. For the each individual selected to the sample, only the variables $S^{(0)}_
i$, $S^{(-1)}_i$, $T_i$ and $T^*_i$ are recorded at the time of the survey, which means the amount of the data from the sample is exiguous compared to the full future purchase histories of 100,000 individuals. To illustrate the effect of the sample size to the accuracy of the estimates, the sample sizes are varied from 100 to 1000.

The model and the chosen prior distributions can be written as follows:
\begin{align*}
 & T_i,T_i^* \sim \textrm{Exp}(\lambda_i), \\
 & \lambda_i \sim \textrm{Gamma}(\gamma,\delta), \\
 & \gamma = m_\lambda^2/v_\lambda, \quad \delta = m_\lambda/v_\lambda, \\
   & m_\lambda,v_\lambda \sim \textrm{Gamma}(2,1),\\
  & S^{(-2)}_i \sim \textrm{Bernoulli}\left( \frac{1-q_i}{2-q_i-p_i} \right), \\
  & S^{(-1)}_i \sim \textrm{Bernoulli}\left( p_i S^{(-2)}_i+(1-q_i)(1-S^{(-2)}_i) \right), \\
  & S^{(0)}_i \sim \textrm{Bernoulli}\left( p_i S^{(-1)}_i+(1-q_i)(1-S^{(-1)}_i) \right), \\
  & p_i \sim \textrm{Beta}(\alpha_p,\beta_p), \quad q_i \sim \textrm{Beta}(\alpha_q,\beta_q), \\
  & \alpha_p = k_p m_p,  \quad \beta_p = k_p (1-m_p),\\
  & \alpha_q = k_q m_q,   \quad \alpha_q = k_q (1-m_q),\\
  & m_p \sim \textrm{Uniform}(0,1), \quad m_q \sim \textrm{Uniform}(0,1), \\
  & k_p \sim \textrm{Gamma}(10,1), \quad k_q \sim \textrm{Gamma}(10,1).
 \\
\end{align*}

For parameters $\gamma$, $\delta$, $\alpha_p$, $\beta_p$, $\alpha_q$ and $\beta_q$ we use weakly informative prior distributions \citep{Gelman:priordistributions}.  Parameters $\gamma$ and $\delta$ describe the shape and scale of the Gamma distribution where the values for the intensity $\lambda_i$ are drawn. We define $\gamma=m_{\lambda}^2/v_{\lambda}$ and $\delta=m_{\lambda}/v_{\lambda}$ where $m_{\lambda} \sim \textrm{Gamma}(2,1)$ is the mean of the intensity distribution and $v_{\lambda} \sim \textrm{Gamma}(2,1)$ is the variance of the intensity distribution. In other words, the expected mean of the intensity distribution is 2 years and the expected standard deviation of the intensity distribution is 1.4 years but there is a considerable uncertainty on the intensity distribution. With these priors, the 95\% Bayes interval for the purchase intervals in the population is $(10^{-2},8 \times 10^{11})$ indicating that the priors are rather uninformative.

Parameters $\alpha_p$ and $\beta_p$ describe the Beta distribution from where the individual repurchase probabilities are drawn and parameters $\alpha_q$ and $\beta_q$ describe the Beta distribution from where the individual competitor repurchase probabilities are drawn. We define $\alpha_p=k_p m_p$ and $\beta_p=k_p (1-m_p)$ where $m_p \sim \textrm{Uniform}(0,1)$ is the expected average repurchase probability and  $k_p \sim \textrm{Gamma}(10,1)$  controls the variation of the repurchase probabilities in the population. Similarly we define $\alpha_q=k_q m_q$ and $\beta_q=k_q (1-m_q)$ where $m_q \sim \textrm{Uniform}(0,1)$ and  $k_q \sim \textrm{Gamma}(10,1)$. These priors for the expected average repurchase probabilities are uninformative but  the $\textrm{Gamma}(10,1)$ for $k_p$ and $k_q$ makes sure that there is a reasonable variation of repurchase probabilities in the population. With these priors, the 95\% Bayes interval for the repurchase probability $p$ in the population is $(0.001,0.999)$ indicating that the prior is otherwise flat but there are peaks near 0 and 1.

\subsection*{BUGS code}
The analysis is carried out using OpenBUGS 3.2.2 \citep{openbugs}, R \citep{R} and R2OpenBUGS R package \citep{R2OpenBUGS}. The BUGS code for the model is given as:
\begin{footnotesize}
\begin{verbatim}
model
{
  for(i in 1:N)
  {
    lambda[i] ~ dgamma(gammal,deltal)
    p[i] ~ dbeta(alphap,betap)
    q[i] ~ dbeta(alphaq,betaq)
    tau[i] ~ dexp(lambda[i])
    taustar[i] ~ dexp(lambda[i])
    m0[i] <- (1-q[i])/(2-q[i]-p[i])
    S2[i] ~ dbern(m0[i])
    S1prob[i] <- p[i]*S2[i]+(1-q[i])*(1-S2[i])
    S1[i] ~ dbern(S1prob[i])
    S0prob[i] <- p[i]*S1[i]+(1-q[i])*(1-S1[i])
    S0[i] ~ dbern(S0prob[i])
  }
  ml ~ dgamma(2,1)
  vl ~ dgamma(2,1)
  gammal <- ml*ml/(vl+0.00001)
  deltal <- ml/(vl+0.00001)
  mp ~ dunif(0,1)
  mq ~ dunif(0,1)
  kp ~ dgamma(10,1)
  kq ~ dgamma(10,1)
  alphap <- kp*mp
  betap <- kp*(1-mp)
  alphaq <- kq*mq
  betaq <- kq*(1-mq)
}
\end{verbatim}
\end{footnotesize}

\subsection*{Simulation results}
\begin{figure}
  \includegraphics[width=\columnwidth]{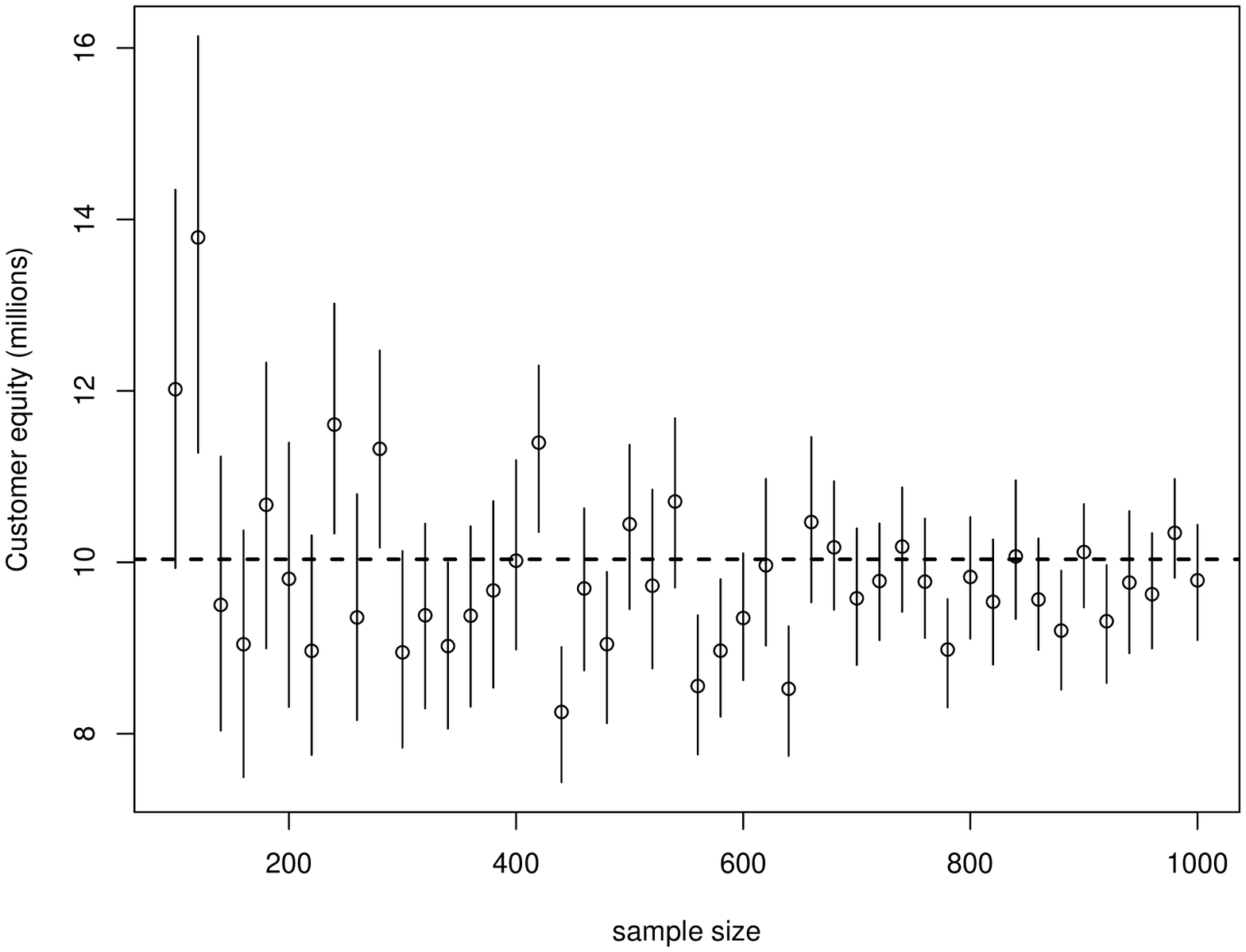}
  \caption{The accuracy of the Bayesian CE estimation as a function of the sample size of the survey. The circles show the mean of the estimated CE posterior distribution and the vertical lines show the posterior range from the 1st decile to the 9th decile. The horizontal line shows the true CE of the population.}\label{fig:simuresults}
\end{figure}

The simulation results are shown in Figure~\ref{fig:simuresults} and in Table~\ref{tab:CLVdistribution}. From Figure~\ref{fig:simuresults}, it can be seen that the estimated posterior distributions are concentrated around the true value of the CE and the systematic bias is small or non-existing. As expected, the variance is smaller for the larger sample sizes. Sample sizes of 800 or more seem to give sufficient accuracy of estimation.

The CLV posterior distributions for the customers of the focal company and the customers of a competitor are presented in Table~\ref{tab:CLVdistribution}. It can be seen that the posteriors estimated from a sample of size 1000 are very similar to the true CLV distribution of the population.
\begin{center}
\begin{longtable}{rcccccc}
\caption{Estimated CLV distributions for the customers of the focal company and the customers of a competitor in the simulation example. \label{tab:CLVdistribution} }\\
\multicolumn{7}{c}{Last purchase with the focal company} \\
N & Min & 1st Qu. & Median & Mean & 3rd Qu. & Max\\
 Population  &  0  &  33.92  &  93.93  &  120.6  &  173.9  &  1122 \\
 1000  &  0  &  35.70  &  93.65  &  118.1  &  169.8  &  1015 \\
 500  &  0  &  40.94  &  101.1  &  129.2  &  185.1  &  1050 \\
 100  &  0  &  56.66  &  125.4  &  150.2  &  210.7  &  865 \\
 & & & & & & \\
\multicolumn{7}{c}{Last purchase with a competitor} \\
N & Min & 1st Qu. & Median & Mean & 3rd Qu. & Max\\
 Population  &  0  &  11.75  &  62.28  &  90.57  &  135.2  &  1184 \\
 1000  &  0  &  12.64  &  62.81  &  88.11  &  131.8  &  966 \\
 500  &  0  &  12.06  &  64.38  &  92.77  &  139.4  &  907.7 \\
 100  &  0  &  16.67  &  74.64  &  101.2  &  152.5  &  857.6
\end{longtable}
\end{center}

\section*{Appendix 2}
\subsection*{BUGS code for the mobile phone data}
The BUGS code for the mobile phone survey is given as
\begin{footnotesize}
\begin{verbatim}
model
{
  for(i in 1:Nretained)
  {
    interval[i] ~ dgamma(kappa,lambda[i])C(it_min[i],it_max[i])
    log(lambda[i]) <- betaconst+betaprev[prevbrand[i]]+betaagegr[agegr[i]]+
                      betagender[gender[i]]+betaincomegr[incomegr[i]]+
                      betaarea[area[i]]  
    repurchase[i] ~ dbern(p[i])
    logit(p[i]) <- alphaconst+alphaprev[prevbrand[i]]+alphaagegr[agegr[i]]+
                   alphagender[gender[i]]+alphaincomegr[incomegr[i]]+
                   alphaarea[area[i]] 
  }
  for(i in (Nretained+1):(Nretained+Nchurned))
  {
    interval[i] ~ dgamma(kappa,lambda[i])C(it_min[i],it_max[i])   
    log(lambda[i]) <- betaconst+betaprev[prevbrand[i]]+betaagegr[agegr[i]]+
                      betagender[gender[i]]+betaincomegr[incomegr[i]]+
                      betaarea[area[i]]   
    repurchase[i] ~ dbern(p[i])
    logit(p[i]) <- alphaconst+alphaprev[prevbrand[i]]+alphaagegr[agegr[i]]+
                   alphagender[gender[i]]+alphaincomegr[incomegr[i]]+
                   alphaarea[area[i]]  
    for(brand in 1:Nbrands)
      {
         q[i,brand] <- (1-equals(prevbrand[i],brand))*qq[brand,agegr[i]]/
			(sum(qq[,agegr[i]])-qq[prevbrand[i],agegr[i]])
      }
      aquisition[i] ~ dbern(q[i,newbrand[i]])
  }
  for(i in (Nretained+Nchurned+1):N)
  {
    interval[i] ~ dgamma(kappa,lambda[i])C(it_min[i],it_max[i])
    log(lambda[i]) <- betaconst+betaprev[prevbrand[i]]+betaagegr[agegr[i]]+
                      betagender[gender[i]]+betaincomegr[incomegr[i]]+
                      betaarea[area[i]]  
  }
  kappa ~ dgamma(1,1)
  betaconst ~ dnorm(0,0.001)
  alphaconst ~ dnorm(0,0.001)
  for(h in 2:Nbrands)
  {
    betaprev[h] ~ dnorm(0,0.001)
    alphaprev[h] ~ dnorm(0,0.001)
  }
  betaprev[1] <- 0
  alphaprev[1] <- 0
  for(h in 2:Nagegr)
  {
    betaagegr[h] ~ dnorm(0,0.001)
    alphaagegr[h] ~ dnorm(0,0.001)
  }
  betaagegr[1] <- 0
  for(h in 2:Nincomegr)
  {
    betaincomegr[h] ~ dnorm(0,0.001)
  }
  betaincomegr[1] <- 0
  for(h in 2:Narea)
  {
    betaarea[h] ~ dnorm(0,0.001)
  }
  betaarea[1] <- 0
  betagender[2] ~ dnorm(0,0.001)
  betagender[1] <- 0
  alphaagegr[1] <- 0
  for(h in 2:Nincomegr)
  {
    alphaincomegr[h] ~ dnorm(0,0.001)
  }
  alphaincomegr[1] <- 0
  for(h in 2:Narea)
  {
    alphaarea[h] ~ dnorm(0,0.001)
  }
  alphaarea[1] <- 0
  alphagender[2] ~ dnorm(0,0.001)
  alphagender[1] <- 0
  for(h1 in 1:Nbrands)
  {
    for(h2 in 1:Nagegr)
    {
        qq[h1,h2] ~ dbeta(2,2)    
    }
  }
}
\end{verbatim}
\end{footnotesize}

\bibliographystyle{chicago_mod}
\bibliography{clv}

\end{document}